\documentclass[conference]{IEEEtran}
\IEEEoverridecommandlockouts
\usepackage{cite}
\usepackage{amsmath,amssymb,amsfonts}
\usepackage{algorithmic}
\usepackage{lipsum}
\usepackage{graphicx}
\usepackage{subfig}
\usepackage{textcomp}
\usepackage{xcolor}
\usepackage{stfloats}
\usepackage{url}
\def\BibTeX{{\rm B\kern-.05em{\sc i\kern-.025em b}\kern-.08em
    T\kern-.1667em\lower.7ex\hbox{E}\kern-.125emX}}
\begin{document}
\title{Impact of Terrestrial Blockage on the Coverage of Integrated Satellite-Terrestrial Networks
}

\author{\IEEEauthorblockN{Joon-Young Park}      
\IEEEauthorblockA{\textit{School of Electrical Engineering} \\
\textit{Korea University}\\
Seoul, South Korea \\
megaduke7@korea.ac.kr}
\and
\IEEEauthorblockN{Byungju Lim}
\IEEEauthorblockA{\textit{Department of Electronic Engineering} \\
\textit{Pukyong National University}\\
Busan, South Korea \\
limbj@pknu.ac.kr}
\and
\IEEEauthorblockN{Young-Chai Ko}
\IEEEauthorblockA{\textit{School of Electrical Engineering} \\
\textit{Korea University}\\
Seoul, South Korea \\
koyc@korea.ac.kr}
}

\maketitle

\begin{abstract}
The integration of non-terrestrial networks (NTNs) with terrestrial networks (TNs) is an important step toward ubiquitous connectivity in sixth-generation (6G). Despite growing interest, the geometric impact of urban blockages on an integrated satellite–terrestrial network (ISTN) has not been rigorously quantified. In this paper, we develop a stochastic geometry-based analytical framework that incorporates a Boolean blockage model to characterize the downlink coverage probability of the ISTN and to provide insights for blockage-aware system design. Our analysis reveals that blockages affect satellite links in two competing ways: while they attenuate desired signals, they can also act as spatial shields that suppress aggregate interference. Leveraging this observation, we analytically show that satellite–terrestrial integration can enhance coverage probability across diverse environments ranging from open areas to dense urban deployments, offering a resilient and mathematically tractable approach to maintaining connectivity under heterogeneous blockage conditions.
\end{abstract}


\section{Introduction}
One of the major challenges toward sixth-generation (6G) wireless systems is the provision of uninterrupted and reliable connectivity in environments where terrestrial infrastructure is either sparsely deployed or difficult to install due to geographical and economic constraints~\cite{Intro1}. To address this challenge, non-terrestrial networks (NTNs) have emerged as a promising solution by complementing conventional terrestrial networks with aerial and spaceborne platforms. Among various NTN architectures, low Earth orbit (LEO) satellite networks have attracted significant attention as a key enabler of global coverage, owing to their relatively low altitude, which allows for reduced propagation delay, improved link budgets, and higher achievable data rates compared with traditional geostationary satellite systems~\cite{Intro2}.

However, relying on a stand-alone LEO satellite network is inherently limited in handling dense traffic demands, which motivates an integrated satellite–terrestrial architecture that jointly exploits the wide-area coverage of satellites and the high-capacity access of terrestrial networks. While most existing studies on satellite–terrestrial networks have analyzed system performance under environments where the line-of-sight (LoS) link is assumed to be dominant, this assumption does not always hold in practical scenarios, since terrestrial blockages such as skyscrapers can easily obstruct the direct signal path and convert it into a non-line-of-sight (NLoS) link. Therefore, it is reasonable, and indeed necessary, to evaluate the performance of an integrated satellite–terrestrial network (ISTN) while explicitly accounting for the presence of such blockages and the resulting coexistence of LoS and NLoS propagation conditions.

\begin{figure} [!t]
\centering
\includegraphics[width=3in]{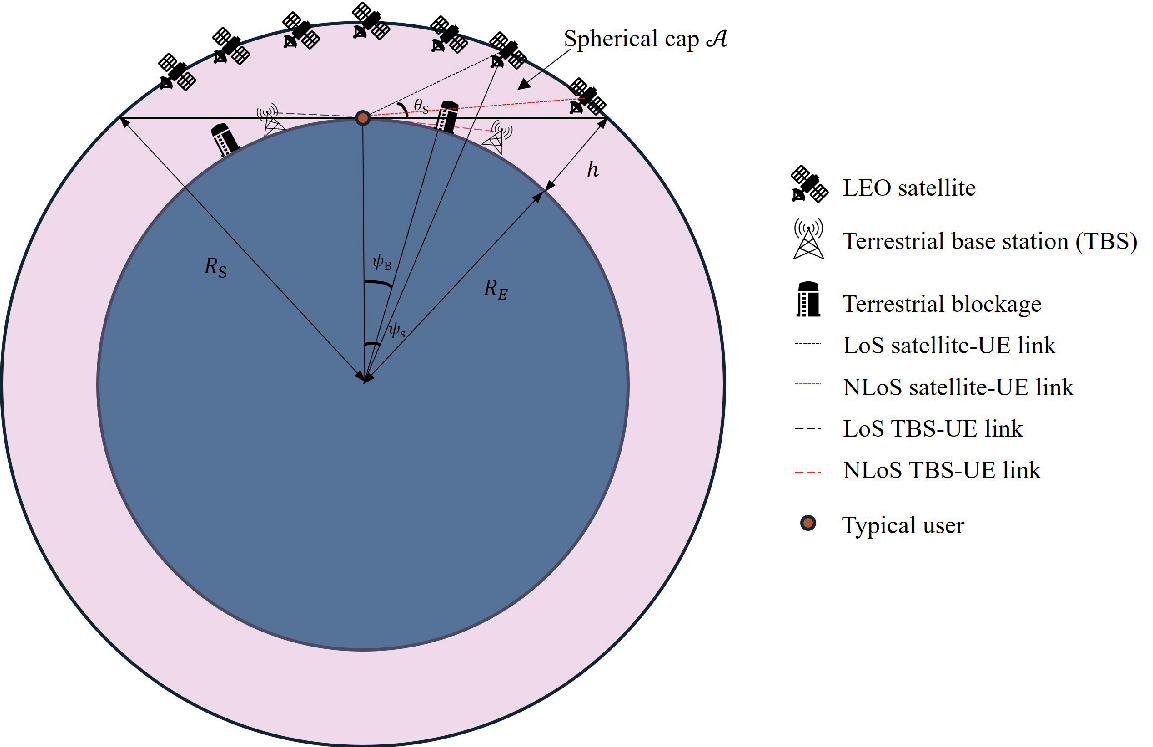}
\caption{Illustration of integrated satellite-terrestrial network (ISTN). The typical user is located at $(0, 0, R_\textnormal{E})$ and the satellites are deployed in the spherical cap $\mathcal{A}$. Terrestrial base stations (TBSs) and the terrestrial blockages are distributed on the Earth while accounting for the Earth's curvature.}
\label{fig1}
\vspace{-5mm}
\end{figure}

Stochastic geometry has been widely adopted as a powerful analytical tool to evaluate the performance of wireless communication networks~\cite{Intro3}. In addition, the effect of the blockage on the coverage of these networks has also been investigated in previous works~\cite{Intro4, Intro5, Intro6}. However, to the best of our knowledge, the impact of terrestrial blockages on the joint performance of the two heterogeneous layers in an ISTN has not yet been rigorously investigated. Motivated by this gap, in this paper we analyze the downlink performance of an ISTN in the presence of terrestrial blockages and quantify how such blockages affect the behavior of this hybrid network. The main contributions of this paper are summarized as follows:
\begin{itemize}
\item We develop a unified stochastic geometry-based analytical framework for downlink coverage analysis of ISTN under a terrestrial blockage-based environment modeled by a Boolean scheme.
\item By explicitly accounting for the curvature of the Earth, we derive a geometric blockage condition for satellite links and obtain a curvature-dependent scaling factor, which leads to a tractable characterization of the LoS satellite process and the satellite-layer coverage probability in the presence of terrestrial blockages.
\item Under a local planar approximation for the terrestrial layer with distance-dependent LoS/NLoS propagation and co-channel interference, we characterize terrestrial downlink coverage and quantify the overall ISTN coverage gain achievable via satellite-terrestrial integration, identifying regimes where integration improves reliability compared to single-network operation.
\end{itemize}
\section{System model}

\subsection{Network Model}

We consider a downlink ISTN as illustrated in Fig.~1. 
The Earth and the satellite shell are embedded in three-dimensional Euclidean space 
$\mathbb{R}^3$, both centered at the origin $\mathbf{0}$, with radii $R_\mathrm{E} \approx 6371~\mathrm{km}$ 
and $R_\mathrm{S}$, respectively. Given a satellite altitude $h$, the radius of the satellite sphere is defined as $R_\mathrm{S} = R_\mathrm{E} + h$. Each satellite location is denoted by $\mathbf{x}$ and represented in spherical coordinates as
\begin{equation}
    \mathbf{x} \triangleq (R_\mathrm{S}, \varphi, \vartheta),~0 \le \varphi \le 2\pi,~0 \le \vartheta \le \pi.
\end{equation}
where $\varphi$ and $\vartheta$ are the azimuth angle and the polar angle, respectively. 

The satellites are modeled as a homogeneous spherical Poisson point process (SPPP) on the sphere with density $\lambda_\mathrm{S}$, and the set of satellite locations is denoted by
$\Phi_\mathrm{S}=\{\mathbf{s}_1, \mathbf{s}_2, \cdots, \mathbf{s}_N\}$. Similarly, downlink users are distributed on the Earth, following a homogeneous SPPP with density $\lambda_\mathrm{U}$, and defined as $\Phi_\mathrm{U}=\{\mathbf{u}_1, \mathbf{u}_2, \cdots, \mathbf{u}_M\}$. Without loss of generality, by Slivnyak's theorem~\cite{Slivnyak}, we assume that the typical user $\mathbf{u}_1$ is located at $(0, 0, R_\mathrm{E})$. We define the spherical cap $\mathcal{A}$ as the portion of the satellite shell above a tangent plane centered at the typical user. Accordingly, the number of visible satellites $N_\mathrm{S}$ follows a Poisson distribution with mean $\mathbb{E}[N_\mathrm{S}]=\lambda_\mathrm{S}|\mathcal{A}|=\lambda_\mathrm{S}2\pi R_\mathrm{S}(R_\mathrm{S}-R_\mathrm{E})$.

For each satellite within the visible spherical cap $\mathcal{A}$, let $\theta_{\mathrm{S}} \in [0, \pi/2]$ denote its elevation angle with respect to the typical user. Consequently, the link distance ranges from $R_{\min} = R_\mathrm{S} - R_\mathrm{E}$ at zenith, to $R_{\max} = \sqrt{R_\mathrm{S}^2 - R_\mathrm{E}^2}$ at horizon. Alternatively, the satellite position can be characterized by the geocentric angle $\psi_\mathrm{S}$, defined as the angle at the Earth's center between the position vectors of the typical user and the satellite. Using simple geometric relations, the elevation angle $\theta_\mathrm{S}$ and the geocentric angle $\psi_\mathrm{S}$ are related as
\begin{equation}
    \theta_{\mathrm{S}}
    = \arctan\left(\frac{R_\mathrm{S} \cos \psi_\mathrm{S} - R_\mathrm{E}}{R_\mathrm{S} \sin \psi_{\mathrm{S}}}\right).
\end{equation}

The terrestrial base stations (TBSs) are distributed on the Earth according to a homogeneous spherical PPP with density $\lambda_\mathrm{T}$, denoted by $\Phi_{\mathrm{T}}=\{\mathbf{t}_1, \mathbf{t}_2, \cdots, \mathbf{t}_Q\}$. Each TBS is mounted on a thin, pole-like structure of height $h_\mathrm{TBS}$ and is therefore not treated as a geometric blockage. Although TBSs are deployed on the Earth, terrestrial links span distances much smaller than $R_\mathrm{E}$; hence, in the performance analysis we approximate the neighborhood of the typical user by the local tangent plane and use a planar PPP model for the terrestrial layer. 

Finally, terrestrial blockages are assumed to be deployed on the Earth, obstructing signal links between the typical user and satellites or TBSs. These blockages are modeled using a Boolean scheme with rectangular footprints~\cite{Intro4}. The blockage centers form a homogeneous PPP, denoted by $\Phi_{\mathrm{B}}$ with intensity $\lambda_\mathrm{B}$. Each blockage is characterized by its footprint dimensions $(L_\mathrm{B}, W_\mathrm{B})$, orientations $\Theta_\mathrm{B}$. The lengths $L_\mathrm{B}$ and the widths $W_\mathrm{B}$ are independent and identically distributed (i.i.d.) positive random variables with probability density functions (PDFs) $f_{L_\mathrm{B}}(\cdot)$ and $f_{W_\mathrm{B}}(\cdot)$, respectively. The orientations $\Theta_\mathrm{B}$ follow a uniform distribution over $[0, 2\pi)$, while the blockage height $h_\mathrm{B}$ follows a Rayleigh distribution.

\subsection{Channel Model}
The received signal power from a TBS $q \in \Phi_{\mathrm{T}}$ to the typical user $\mathbf{u}_1$ is given by
\begin{equation}
   P_{\mathrm{T},q}=P_{t, \mathrm{T}}G_{\mathrm{T,q}}H_{\mathrm{T}, q}\|\mathbf{t}_q-\mathbf{u}_1\|^{-{\alpha}_\mathrm{T}},
\end{equation}
where $P_{t,\mathrm{T}}$ denotes the transmit power of the TBS, $G_{\mathrm{T,q}}$ denotes the effective transmit antenna gain on the link from TBS $q$ to the typical user. We adopt a deterministic two-sectored antenna model in which the serving TBS is perfectly aligned with the typical user and thus provides the main-lobe gain $G_{\mathrm{T,q}}=G_{\mathrm{main,T}}$, while all interfering TBSs are assumed to contribute only through side-lobes toward the typical user, with gain $G_{\mathrm{T,q}}=G_{\mathrm{side,T}}$. The path loss exponent is $\alpha_\mathrm{T}=\alpha_\mathrm{L}$ for LoS links and $\alpha_\mathrm{T}$=$\alpha_\mathrm{N}$ for NLoS links. The small-scale fading power $H_{\mathrm{T}, q}$ depends on the link condition: for LoS links, $H_{\mathrm{T}, q}$ follows Nakagami-$m$ fading, \textit{i.e.,} $H_{\mathrm{T},q}\sim\mathrm{Gamma}(m_L,1/m_L)$;
for NLoS links, $H_{\mathrm{T},q}$ follows Rayleigh fading in power, \textit{i.e.,} $H_{\mathrm{T},q}\sim\mathrm{Exp}(1)$ with unit mean. Similarly, the received signal power from a satellite $n\in\Phi_\mathrm{S}$ is expressed as
\begin{equation}
    P_{\mathrm{S},n}=P_{t,\mathrm{S}}G_{\mathrm{S,n}}H_{\mathrm{S,n}}\|\mathbf{s}_n-\mathbf{u}_1\|^{-\alpha_\mathrm{S}},
\end{equation}
where $P_{t,\mathrm{S}}$ denotes the satellite transmit power, $G_{\mathrm{S,n}}$ denotes the effective transmit antenna gain on the link from satellite $n$ to the typical user, and $\alpha_\mathrm{S}$ is the path-loss exponent for the satellite link. Similar to the terrestrial case, $G_{\mathrm{S,n}}=G_{\mathrm{main,S}}$ for the serving satellite and $G_{\mathrm{S,n}}=G_{\mathrm{side,S}}$ for interfering satellites. Here, we consider a direct-to-cell operation. For satellite links, we adopt a two-state availability model, where a link is either LoS or blocked by terrestrial blockages. Blocked links are treated as being in outage and thus contribute neither desired signal nor interference. This hard-blockage assumption is adopted for analytical tractability. Moreover, the terrestrial and satellite downlinks operate on orthogonal carriers, and thus cross-tier interference is neglected throughout the paper.

Conditioned on being LoS, the small-scale fading of satellite links is modeled by the shadowed-Rician fading model, which effectively captures the combined effects of multipath fading and random shadowing on the dominant LoS component in satellite communications~\cite{Jung}. Although this model provides a comprehensive characterization, its probability density function involves complicated hypergeometric functions, rendering direct analytical treatment intractable. It has been shown, however, that the shadowed-Rician fading power can be accurately approximated by a Gamma random variable~\cite{SR}. We employ this approximation in our analysis, as it provides both high accuracy and analytical tractability.

\section{Performance analysis under the blockages}
\subsection{Coverage Analysis of the Satellite Network Under Terrestrial Blockages}
Since the LEO satellites operate at high altitudes relative to the Earth, the LoS condition between a satellite and a ground user is strongly influenced by the curvature of the Earth. Unlike conventional flat-Earth models, where blockage is typically determined solely by the elevation angle and blockage height, the Earth's curvature alters the effective relative height of terrestrial blockages along the propagation path. In particular, a blockage located farther from the user appears lower with respect to the local tangent plane,  which reduces its ability to obstruct the satellite link.

To rigorously capture this geometric effect, we first derive the exact condition under which a terrestrial blockage obstructs the LoS path between a satellite and the typical user. Based on this condition, we characterize the LoS probability of satellite links in the presence of randomly distributed blockages and subsequently analyze the resulting coverage probability of the satellite network.

\textit{Lemma 1: The average blockage probability scaling factor $\eta(\psi_\mathrm{S})$ for the LoS link between the typical user and the satellite with the geocentric angle $\psi_\mathrm{S}$ is given by}
\begin{equation}\label{Lemma1}
\eta(\psi_\mathrm{S})=
\begin{cases}
\frac{1}{\psi_\mathrm{S}}\int_{0}^{\psi_\mathrm{S}}
\exp\!\left(-\frac{\mathcal{H}_{\mathrm{th}}(\psi_\mathrm{B}, \psi_\mathrm{S})^2}{2\sigma_\mathrm{B}^2}\right)\,d\psi_\mathrm{B}, & \psi_\mathrm{S}>0,\\[6pt]
1, & \psi_\mathrm{S}=0.
\end{cases}
\end{equation}
\textit{where $\psi_\mathrm{B}$ denotes the geocentric angle of the blockage, $\sigma_\mathrm{B}$ is the scale parameter of the Rayleigh distribution modeling the blockage height, and $\{\mathcal{H}_{\mathrm{th}}(\psi_\mathrm{B}, \psi_\mathrm{S})\}$ represents the minimum blockage height required to obstruct the LoS link}.

\quad\textit{Proof:~} Consider a terrestrial blockage with random height $h_\mathrm{B}$ located at an intermediate geocentric angle $\psi_\mathrm{B}$, where $0\leq\psi_\mathrm{B}\leq\psi_\mathrm{S}$ along the orthodromic path between the typical user and the satellite. Due to the curvature of the Earth, the blockage is tilted relative to the local tangent plane at the user by an angle $\psi_\mathrm{B}$. To determine whether the blockage obstructs the LoS link, we compare its effective vertical height, projected onto the user's tangent plane, with the height of the LoS ray at the same horizontal location. The projected vertical height of the blockage is given by 
\begin{equation}
z_\mathrm{B}=(R_\mathrm{E}+h_\mathrm{B})\cos(\psi_\mathrm{B})-R_\mathrm{E},
\end{equation}
while the height of the LoS ray evaluated at the horizontal coordinate of the blockage top point, $x_T=(R_\mathrm{E}+h_\mathrm{B})\sin\psi_\mathrm{B}$, is 
\begin{equation}
z_\mathrm{link}=x_T\mathcal{T}(\psi_\mathrm{S}),
\end{equation}
where 
\begin{equation}
\mathcal{T}(\psi_\mathrm{S})=\frac{R_\mathrm{S}\cos(\psi_\mathrm{S})-R_\mathrm{E}}{R_\mathrm{S}\sin(\psi_\mathrm{S})}.
\end{equation} The blockage obstructs the LoS link if and only if its projected height exceeds the ray height, \textit{i.e.}, $z_\mathrm{B}>z_\mathrm{link}$. Substituting the above expressions yields the condition:
\begin{equation}
(R_\mathrm{E} + h_\mathrm{B})\cos\psi_\mathrm{B}- R_\mathrm{E} > (R_\mathrm{E} + h_\mathrm{B})\sin\psi_\mathrm{B} \cdot \mathcal{T}(\psi_\mathrm{S}).
\end{equation}
Rearranging this inequality with respect to the blockage height $h_\mathrm{B}$, we obtain the minimum height required to block the LoS link at location $\psi_\mathrm{B}$, denoted as the threshold height $\mathcal{H}_{\mathrm{th}}(\psi_\mathrm{B}, \psi_\mathrm{S})$:
\begin{equation}
h_\mathrm{B} > R_\mathrm{E} \left( \frac{1}{\cos\psi_\mathrm{B} - \sin\psi_\mathrm{B} \cdot \mathcal{T}(\psi_\mathrm{S})} - 1 \right) \triangleq \mathcal{H}_{\mathrm{th}}(\psi_\mathrm{B}, \psi_\mathrm{S}).
\end{equation}
Following the Boolean blockage model, the blockage centers form a Poisson point process, and the blockage location $\psi_\mathrm{B}$ is uniformly distributed along the path $[0, \psi_\mathrm{S}]$. Therefore, the average blockage probability scaling factor is obtained by averaging the probability that the blockage height exceeds the threshold height over the entire path, \textit{i.e.},
\begin{align}
\eta(\psi_\mathrm{S}) &= \mathbb{E}_{\Psi_\mathrm{B}} \left[ \mathbb{P}(h_\mathrm{B} > \mathcal{H}_{\mathrm{th}}(\psi_\mathrm{B}, \psi_\mathrm{S})) \right] \nonumber\\
&= \frac{1}{\psi_\mathrm{S}} \int_{0}^{\psi_\mathrm{S}} \bar{F}_{h_\mathrm{B}}(\mathcal{H}_{\mathrm{th}}(\psi_\mathrm{B}, \psi_\mathrm{S})) d\psi_\mathrm{B},
\end{align}
where $\bar{F}_{h_\mathrm{B}}$ denotes the complementary cumulative distribution function (CCDF) of the blockage height $h_\mathrm{B}$. Assuming that the blockage height follows a Rayleigh distribution with scale parameter $\sigma_\mathrm{B}$, the expression in (\ref{Lemma1}) is obtained, which completes the proof.\hfill $\blacksquare$

The scaling factor $\eta(\psi_\mathrm{S})$ quantifies the reduction in blockage effectiveness induced by the Earth's curvature. Specifically, as the geocentric angle increases, terrestrial blockages appear lower relative to the user's local tangent plane, which decreases their likelihood of obstructing the satellite link. By incorporating this factor, the LoS probability between the typical user and a satellite can be accurately characterized as a function of the satellite's geocentric angle.

In prior work~\cite{Intro4}, the number of blockages intersecting a link of length $R$ in planar networks has been modeled as a Poisson random variable with mean $\mathbb{E}[K]=\beta R +p$. Here, $\beta=2\lambda_\mathrm{B}(\mathbb{E}[L_\mathrm{B}]+\mathbb{E}[W_\mathrm{B}])/\pi$ and $p=\lambda_\mathrm{B}\mathbb{E}[L_\mathrm{B}]\mathbb{E}[W_\mathrm{B}]$. To account for Earth curvature in the satellite scenario, we extend this result by replacing the planar link length with the corresponding orthodromic distance on the Earth's surface and by incorporating the curvature-dependent scaling factor derived in Lemma 1.

\textit{Lemma 2: For a satellite $n$ whose geocentric angle to the typical user is $\psi$,
the LoS probability of the link $\mathbf{s}_n-\mathbf{u}_1$ is given by}
\begin{align}\label{Lemma2}
P_{\mathrm{LoS}}^{\mathrm{Sat}}(\psi)
=
\begin{cases}
e^{-\eta(\psi)(\beta R_\mathrm{E} \psi + p)}, & 0\le \psi \le \psi_{\max}\\
0, & \text{otherwise}.
\end{cases}
\end{align}
\textit{where $\beta$ and $p$ are the blockage parameters defined in~\cite{Intro4}, and $\psi_{\mathrm{max}}=\arccos\left(\frac{R_\mathrm{E}}{R_\mathrm{S}}\right)$}.

\quad\textit{Proof:~} In planar blockage models, the number of blockages intersecting a link of length $R$ is commonly modeled as a Poisson random variable with mean $\mathbb{E}[K]=\beta R + p$~\cite{Intro4}. In the considered satellite scenario, the orthodromic projection of the satellite-user link with geocentric angle $\psi$ onto the Earth's surface has length $R_\mathrm{E}\psi$. Accordingly, the number of potential blockages along this path follows
\begin{equation}
K_{\mathrm{potential}}\sim\mathrm{Poisson}(\beta R_\mathrm{E} \psi+p).
\end{equation}
Due to the Earth's curvature, not all blockages located along this path are effective in obstructing the LoS link. By incorporating the curvature-induced blockage effectiveness through the scaling factor $\eta(\psi)$ derived in Lemma 1, the number of effective blockages is modeled as
\begin{equation}
K_{\mathrm{eff}}\sim\mathrm{Poisson}\left(\eta(\psi)(\beta R_\mathrm{E} \psi+p)\right).
\end{equation} 
Using the void probability of the Poisson point process, the LoS probability is obtained as $P_{\mathrm{LoS}}^{\mathrm{Sat}}(\psi)=\mathbb{P}[K_\mathrm{eff}=0]=e^{-\left(\eta(\psi)(\beta R_\mathrm{E} \psi+p)\right)}$. Finally, satellites with geocentric angles larger than $\psi_{\mathrm{max}}$ are not visible to the typical user and thus yield zero LoS probability, which completes the proof.\hfill $\blacksquare$

Based on Lemma 2, the set of LoS satellites can be modeled as a thinned SPPP, whose intensity is a function of the geocentric angle. In the following lemma, we characterize the distribution of the geocentric angle of the nearest LoS satellite associated with the typical user.

\textit{Lemma 3: Let $\Psi_\mathrm{S}$ denote the geocentric angle of the nearest LoS satellite as observed from the typical user $\mathbf{u}_1$, conditioned on the event that at least one LoS satellite exists within the visible region $[0, \psi_\mathrm{max}].$ Under this condition, the PDF of $\Psi_\mathrm{S}$ is given by (\ref{Lemma3}).}

\quad\textit{Proof (sketch):~}
LoS satellites form an inhomogeneous PPP on the visible cap, obtained by thinning the SPPP with the LoS probability $P_{\mathrm{LoS}}^{\mathrm{Sat}}(\psi)$. The surface area of a spherical ring between $\psi$ and $\psi+d\psi$ is
$dA(\psi)=2\pi R_\mathrm{S}^2\sin(\psi)\,d\psi$; hence the mean number of LoS satellites in $[0,\psi]$ is $\mu(\psi)=\int_0^\psi \lambda_\mathrm{S}P_{\mathrm{LoS}}^{\mathrm{Sat}}(x)\,dA(x)
=\int_0^\psi 2\pi R_\mathrm{S}^2\sin(x)\lambda_\mathrm{S}P_{\mathrm{LoS}}^{\mathrm{Sat}}(x)\,dx$. Therefore, by the void probability, $\mathbb{P}(\Psi_\mathrm{S}>\psi)=\exp(-\mu(\psi))$. Conditioning on $N_{\mathrm{L}}(\psi_{\max})\ge 1$ and differentiating the conditional CDF yields \eqref{Lemma3}. \hfill $\blacksquare$

Conditioned on the event that the typical user is associated with the nearest LoS satellite at geocentric angle $\psi_\mathrm{S}$, the remaining LoS satellites lie in the region $(\psi_\mathrm{S}, \psi_\mathrm{max}]$ and form an inhomogeneous Poisson process with intensity $2\pi R_\mathrm{S}^2 \sin(\psi)\lambda_\mathrm{S}P_\mathrm{LoS}^{\mathrm{Sat}}(\psi)$. Under this condition, the aggregate interference power from the satellite network can be characterized.

\textit{Lemma 4: Conditioned on the event that the typical user is associated with the nearest LoS satellite at geocentric angle $\psi_\mathrm{S}$, the Laplace transform of the aggregate interference from remaining LoS satellites is given by (\ref{Lemma4}). Here, we define}
\begin{equation}
m_s=\frac{m(2b_0+\Omega)^2}{4mb_0^2+4m b_0 \Omega + \Omega^2}
\end{equation} \textit{and}
\begin{equation}
\beta_s=\frac{4mb_0^2 +4m b_0 \Omega + \Omega^2}{m(2b_0+\Omega)}
\end{equation}
\textit{where $b_0,m$ and $\Omega$ are the parameters of the shadowed-Rician fading, and $d(\psi)=\sqrt{R_\mathrm{E}^2+R_\mathrm{S}^2-2R_\mathrm{E} R_\mathrm{S} \cos(\psi)}$}.

\begin{figure*}[t!]
\centering
\normalsize
\begin{align}
&f_{\Psi_\mathrm{S}}(\psi_\mathrm{S})=2\pi R_\mathrm{S}^2\sin(\psi_\mathrm{S})\lambda_\mathrm{S} P_{\mathrm{LoS}}^{\mathrm{Sat}}(\psi_\mathrm{S})\frac{\exp\left(-\int_{0}^{\psi_\mathrm{S}}{2\pi R_\mathrm{S}^2\sin(x)\lambda_\mathrm{S} P_{\mathrm{LoS}}^{\mathrm{Sat}}(x)~dx}\right)}{1-\exp\left(-\int_{0}^{\psi_\mathrm{max}}{2\pi R_\mathrm{S}^2\sin(x)\lambda_\mathrm{S} P_{\mathrm{LoS}}^{\mathrm{Sat}}(x)~dx}\right)}\label{Lemma3}\\
&\mathcal{L}_{I_{\mathrm{Sat}}}(s|\psi_\mathrm{S}) = \exp \Bigg( - \int_{\psi_\mathrm{S}}^{\psi_{\mathrm{max}}} \left[ 1 - \left( 1 + s \beta_s P_{t,\mathrm{S}} G_{\mathrm{side,S}}\,(d(\psi))^{-\alpha_\mathrm{S}} \right)^{-m_s} \right] 2\pi R_\mathrm{S}^2 \sin(\psi) \lambda_\mathrm{S} P_{\mathrm{LoS}}^{\mathrm{Sat}}(\psi) \, d\psi \Bigg)\label{Lemma4} \\
&P_{\mathrm{Cov}}^{\mathrm{Sat}}= \mathbb{P}~\!\big(\mathrm{SINR}_{\mathrm{Sat}}>\tau,\; N_{\mathrm{L}}(\psi_{\max})\ge 1\big)=\mathbb{P}~\!\big(N_{\mathrm{L}}(\psi_{\max})\ge 1\big)
\,\mathbb{E}\!\left[\mathbb{P}\!\left(\mathrm{SINR}_{\mathrm{Sat}}>\tau \mid \Psi_{\mathrm{S}}\right)\Big|\, N_{\mathrm{L}}(\psi_{\max})\ge 1\right]\label{Proposition1}\\
&\mathcal{L}_{I_{\mathrm{Ter}}}(s \mid r_0) = \exp\left(-2\pi \lambda_\mathrm{T} \int_{r_0}^{\infty} \left[1 - \left(1 + \frac{s P_{t, \mathrm{T}}G_{\mathrm{side,T}}x^{-\alpha_\mathrm{L}}}{m_L}\right)^{-m_L}\right] P_\mathrm{LoS}^{\mathrm{Ter}}(x) x \, dx\right) \nonumber \\
&\quad\quad\quad\quad\,\,\quad \times \exp\left(-2\pi \lambda_\mathrm{T} \int_{r_0}^{\infty} \left[1 - \frac{1}{1 + s P_{t,\mathrm{T}}G_{\mathrm{side,T}}x^{-\alpha_\mathrm{N}}}\right] P_\mathrm{NLoS}^{\mathrm{Ter}}(x) x \, dx\right)\label{Lemma5}\\
&P_{\mathrm{Cov}}^{\mathrm{Ter}} = \int_{0}^{\infty} f_{R_0}(r_0) \Bigg[  P_{\mathrm{LoS}}^{\mathrm{Ter}}(r_0) \sum_{k=0}^{m_L-1} \frac{1}{k!} \left(-\frac{d}{ds}\right)^k \left(\mathcal{L}_{I_\mathrm{Ter}}(s \mid r_0) e^{-s N_{0,\mathrm{T}}}\right)\Bigg|_{s=s_\mathrm{L}}+ (1 - P_\mathrm{LoS}^{\mathrm{Ter}}(r_0)) e^{-s_\mathrm{N} N_{0,\mathrm{T}}} \mathcal{L}_{I_\mathrm{Ter}}(s_\mathrm{N}\mid r_0) \Bigg] \, dr_0  \nonumber \\
\label{Proposition2}
\end{align}
\hrulefill
\vspace*{4pt}
\end{figure*}

\quad\textit{Proof (sketch):~} Conditioned on the nearest LoS satellite at $\psi_{\mathrm{S}}$, the interfering satellites form an inhomogeneous SPPP on $(\psi_{\mathrm{S}},\psi_{\max}]$ with intensity $2\pi R_{\mathrm{S}}^{2}\sin(\psi)\lambda_{\mathrm{S}}P_{\mathrm{LoS}}^{\mathrm{Sat}}(\psi)$. By applying the PGFL of the PPP to the aggregate interference and using the Gamma approximation for the fading power $\mathbb{E}_{H}[e^{-tH}]=(1+t\beta_s)^{-m_s}$, the Laplace transform is directly obtained as (\ref{Lemma4}). \hfill $\blacksquare$

\textit{Proposition 1.}
\textit{The downlink coverage probability of the satellite network is given by}
\begin{align}
P_{\mathrm{Cov}}^{\mathrm{Sat}}
&= \mathbb{P}\!\big(N_{\mathrm{L}}(\psi_{\max})\ge 1\big)
\int_{0}^{\psi_{\max}} C_{\mathrm{Sat}}(\psi,\tau)\, f_{\Psi_{\mathrm{S}}}(\psi)\, d\psi \nonumber\\
&=\left(1-e^{-\mu(\psi_{\max})}\right)\int_{0}^{\psi_{\max}} C_{\mathrm{Sat}}(\psi,\tau)\, f_{\Psi_{\mathrm{S}}}(\psi)\, d\psi,
\end{align}
\textit{where $f_{\Psi_{\mathrm{S}}}(\cdot)$ is the PDF of the nearest LoS satellite geocentric angle conditioned on $N_{\mathrm{L}}(\psi_{\max})\ge 1$ given in Lemma~3, $\tau$ denotes the target SINR threshold and the conditional coverage probability $C_{\mathrm{Sat}}(\psi_\mathrm{S}, \tau)$ is expressed as}
\begin{align}
C_{\mathrm{Sat}}(\psi_\mathrm{S}, \tau)
=\sum_{k=0}^{n_s}\frac{(-s_\mathrm{Sat})^k}{k!}
\left.\frac{d^k}{ds^k}\left(e^{-sN_{0,\mathrm{S}}}\mathcal{L}_{I_{\mathrm{Sat}}}(s|\psi_\mathrm{S})\right)\right|_{s=s_{\mathrm{Sat}}},
\end{align}
\textit{where $n_s=\tilde{m}_s- 1$ and $s_\mathrm{Sat}=\frac{\tau (d(\psi_\mathrm{S}))^{\alpha_\mathrm{S}}}{P_{t,\mathrm{S}}G_{\mathrm{main,S}}\beta_s}$.}

\quad\textit{Proof:~}
Since the typical user is served by the nearest LoS satellite, coverage occurs only if there exists at least one LoS satellite in the visible region and the resulting SINR exceeds $\tau$. Hence, we can obtain (\ref{Proposition1}), and using the conditional PDF $f_{\Psi_{\mathrm{S}}}(\cdot)$ in Lemma~3 yields the stated integral form.
Given $\Psi_{\mathrm{S}}=\psi$, the conditional coverage probability is
\begin{align}
C_{\mathrm{Sat}}(\psi,\tau)
=\mathbb{P}\!\left(
\frac{P_{t,\mathrm{S}}G_{\mathrm{main,S}}H_{\mathrm{S},0}d(\psi)^{-\alpha_{\mathrm{S}}}}
{I_{\mathrm{Sat}}+N_{0,\mathrm{S}}}>\tau
\;\Big|\;\Psi_{\mathrm{S}}=\psi
\right),
\end{align}
which can be evaluated by the Gamma approximation of the shadowed-Rician fading and the
Laplace transform $\mathcal{L}_{I_{\mathrm{Sat}}}(s\mid \psi)$ in Lemma~4. Here, $H_{\mathrm{S},0}$ denotes the fading power of the serving satellite link and $N_{0,\mathrm{S}}$ is the thermal noise power in the satellite network. Note that we approximate $m_s$ by the nearest integer $\tilde{m}_s$ to keep a closed-form finite-sum expression, which is known to yield negligible error. \hfill $\blacksquare$
\subsection{Coverage Analysis of the Terrestrial Network Under Terrestrial Blockages}
In the terrestrial network, we assume that the blockage environment is shared with the satellite network. Accordingly, the same blockage parameters are used to characterize the blockages affecting both terrestrial and satellite links. The typical user is associated with its nearest TBS, which serves as the desired transmitter, while all other TBSs act as sources of interference. 

In~\cite{Intro4}, the analysis is conducted under the assumption that the radio signals are impenetrable through blockages, and only LoS links are considered effective. In sub-6GHz systems, however, NLoS components can also contribute non-negligibly to the received signal power. To account for this effect, we explicitly model both LoS and NLoS propagation by adopting different path-loss exponents for the two link conditions. Unlike satellite links, terrestrial links typically operate over relatively short distances. As a result, the curvature of the Earth has a negligible impact on terrestrial propagation, and it is reasonable to conduct the terrestrial network analysis using a two-dimensional planar model.

\textit{Corollary 1.1} The PDF of the distance $r_0$ between the typical user and its serving TBS is given by $f_{R_0}(r_0)=2\pi\lambda_\mathrm{T}r_0\exp(-\pi\lambda_\mathrm{T}r_0^2)$ for $r_0 \ge 0$.

\textit{Corollary 1.2} The LoS probability of a terrestrial network is $P_{\mathrm{LoS}}^{\mathrm{Ter}}(r)=e^{-\eta_\mathrm{Ter}(\beta r + p)}$, as derived in~\cite{Intro4}. Consequently, the NLoS probability is 
\begin{equation}
P_{\mathrm{NLoS}}^{\mathrm{Ter}}(r)=1-e^{-\eta_\mathrm{Ter}(\beta r + p)}.
\end{equation}
where $\eta_\mathrm{Ter}
=\sqrt{\frac{\pi}{2}}\frac{\sigma_\mathrm{B}}{h_\mathrm{TBS}}
\operatorname{erf}\!\left(\frac{h_\mathrm{TBS}}{\sqrt{2}\,\sigma_\mathrm{B}}\right)$, $h_\mathrm{TBS}$ is the height of the TBS, and $\mathrm{erf}$ is the error function.

\textit{Lemma 5.} \textit{Conditioned on the serving distance $r_0$, the Laplace transform of the aggregate terrestrial interference is given by (\ref{Lemma5})}.

\quad\textit{Proof:~} The aggregate interference experienced by the typical user from the terrestrial network can be expressed as
\begin{align}
I_\mathrm{Ter} &= \sum_{i\in\Phi_\mathrm{T,L}\cap \mathcal{B}_{r_0}^{c}}{P_{t,\mathrm{T}}G_{\mathrm{side,T}}H_{\mathrm{T,L},i}{r_i}^{-\alpha_{\mathrm{L}}}} \nonumber \\
&\quad + \sum_{j\in\Phi_\mathrm{T,N}\cap \mathcal{B}_{r_0}^{c}}{P_{t,\mathrm{T}}G_{\mathrm{side,T}}H_{\mathrm{T,N},j}{r_j}^{-\alpha_{\mathrm{N}}}},
\end{align}
where $\mathcal{B}_{r_0}^c$ denotes the region outside a disk of radius $r_0$.
The sets $\Phi_{\mathrm{T,L}}$ and $\Phi_{\mathrm{T,N}}$ represent the LoS and NLoS interferer point processes obtained via distance-dependent thinning with probabilities $P_{\mathrm{LoS}}^{\mathrm{Ter}}(x)$ and $1-P_{\mathrm{LoS}}^{\mathrm{Ter}}(x)$, respectively, as defined in Section~II. $H_{\mathrm{T,L},i}$ and $H_{\mathrm{T,N},j}$ denote the small-scale fading power on the LoS and NLoS interfering link, independent across links and independent of $\Phi_\mathrm{T}$. Using the independence of fading and the PGFL of the PPP, together with the moment generating functions of the fading power distributions specified in Section~II, yields (\ref{Lemma5}), which completes the proof.\hfill $\blacksquare$

\textit{Proposition 2.} \textit{The coverage probability of the terrestrial network is given by (\ref{Proposition2}), where $s_\mathrm{L}=\frac{m_L  r_0^{\alpha_\mathrm{L}}\tau}{P_{t,\mathrm{T}}G_{\mathrm{main},\mathrm{T}}}$ and $s_\mathrm{N}=\frac{r_0^{\alpha_\mathrm{N}}\tau}{P_{t,\mathrm{T}}G_{\mathrm{main},\mathrm{T}}}$}, and $N_{0,\mathrm{T}}$ is the thermal noise power in the terrestrial network.

\quad\textit{Proof:~} The proof follows a procedure similar to that of Proposition 1 and is therefore omitted for brevity.\hfill $\blacksquare$

\subsection{Coverage Probability of ISTN}
In the considered ISTN architecture, the terrestrial network serves as the primary access layer, while the satellite network operates as a complementary layer to enhance reliability. Specifically, the typical user is first served by the terrestrial network; only when the downlink SINR from the terrestrial network falls below the target threshold does the user associate with the satellite network. Under this association policy, the overall coverage probability of the ISTN can be expressed as follows.

\textit{Proposition 3.} \textit{Under a commonly used independence approximation between the terrestrial and satellite coverage events, the coverage probability of the ISTN is approximated as}
\begin{align}
P_\mathrm{Cov}^\mathrm{ISTN} \approx 1-(1-P_\mathrm{Cov}^\mathrm{Ter})(1-P_\mathrm{Cov}^\mathrm{Sat})
\end{align}
\textit{where $P_\mathrm{Cov}^\mathrm{Sat}$ and $P_\mathrm{Cov}^\mathrm{Ter}$ are given in Proposition 1 and 2, respectively.}

\quad\textit{Proof:~}
Coverage occurs if at least one of the two links meets the SINR target.
Using the common independence approximation between the terrestrial and satellite coverage events,
\[
\mathbb{P}(\mathrm{SINR}_\mathrm{Ter}\le\tau,\ \mathrm{SINR}_\mathrm{Sat}\le\tau)
\approx (1-P_\mathrm{Cov}^\mathrm{Ter})(1-P_\mathrm{Cov}^\mathrm{Sat}),
\]
which yields $P_\mathrm{Cov}^\mathrm{ISTN}\approx 1-(1-P_\mathrm{Cov}^\mathrm{Ter})(1-P_\mathrm{Cov}^\mathrm{Sat})$. \hfill $\blacksquare$

\quad\textit{Remark 1:~}Proposition~3 assumes an independence between terrestrial and satellite coverage events. This neglects the coupling induced by the common blockage realization, which typically yields positive correlation in LoS probability. Nevertheless, the independence approximation yields a compact system-level expression and provides a baseline for quantifying the coverage gain achieved by ISTN.

\section{Simulation Results}
In this section, we validate the proposed analytical framework through simulations to examine the impact of blockages on the downlink coverage probability of the ISTN. The simulation parameters are summarized as follows. The radius of the Earth $R_\mathrm{E}$ is set to $6,371~\mathrm{km}$ and the satellite altitude is $h=550~\mathrm{km}$. The transmit powers of the satellite and the TBS are $P_{t,\mathrm{S}}=50~\mathrm{dBm}$ and $P_{t,\mathrm{T}}=40~\mathrm{dBm}$, respectively. The Nakagami fading parameter is set to $m_L=2$, and the path-loss exponents for the satellite, LoS TBS, and NLoS TBS are chosen as 2, 2.5, 4, respectively. The height of TBS $h_\mathrm{TBS}$ is set to 35~$\mathrm{m}$. The thermal noise powers are $N_{0,\mathrm{T}}=-94\mathrm{dBm}$ and $N_{0,\mathrm{S}}=-107\mathrm{dBm}$, corresponding to bandwidths $W_\mathrm{T}=100\mathrm{MHz}$ and $W_\mathrm{S}=5\mathrm{MHz}$. For the satellite link, we adopt a shadowed-Rician fading model under the infrequent light shadowing (ILS) condition, with parameters $(m, b_0,\Omega)=(19,\,0.158,\,1.29)$. The carrier frequency is $3.5~\mathrm{GHz}$ for the TBS and $1.99~\mathrm{GHz}$ for the satellite. The transmit antenna gain of the desired satellite link is set to $38~\mathrm{dBi}$, while that of interfering satellite links is $28~\mathrm{dBi}$~\cite{SpaceX}. For TBSs, the transmit antenna gain is set to 0 dBi. The users are assumed to be equipped with omnidirectional antennas, and thus the receive antenna gain is set to $0~\mathrm{dBi}$.
\begin{figure}[!t]
\centering

\begin{minipage}[t]{0.49\linewidth}
\centering
\subfloat[]{
    \includegraphics[width=\linewidth]{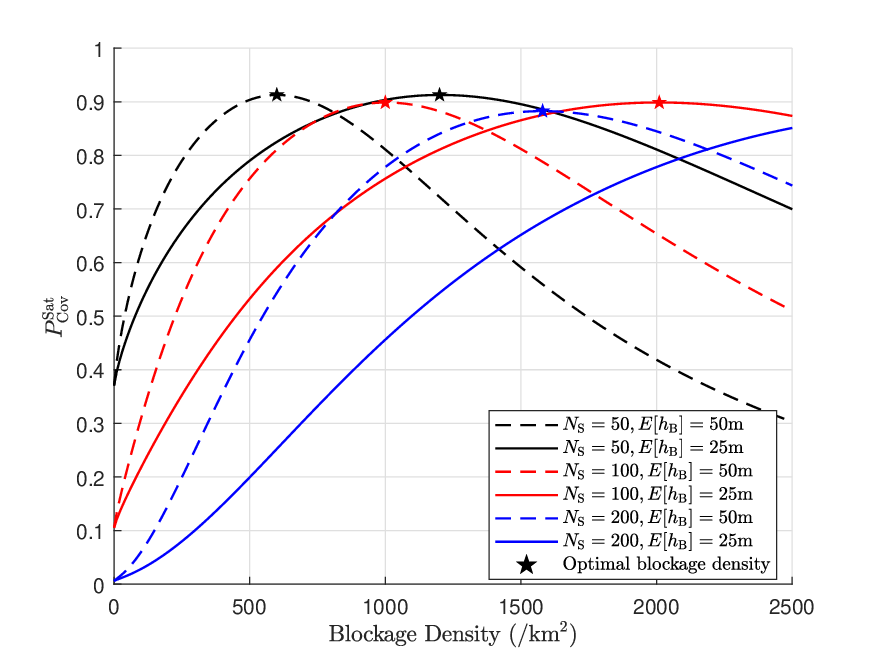}
    \label{fig2a}
}
\end{minipage}
\hfill
\begin{minipage}[t]{0.49\linewidth}
\centering
\subfloat[]{
    \includegraphics[width=\linewidth]{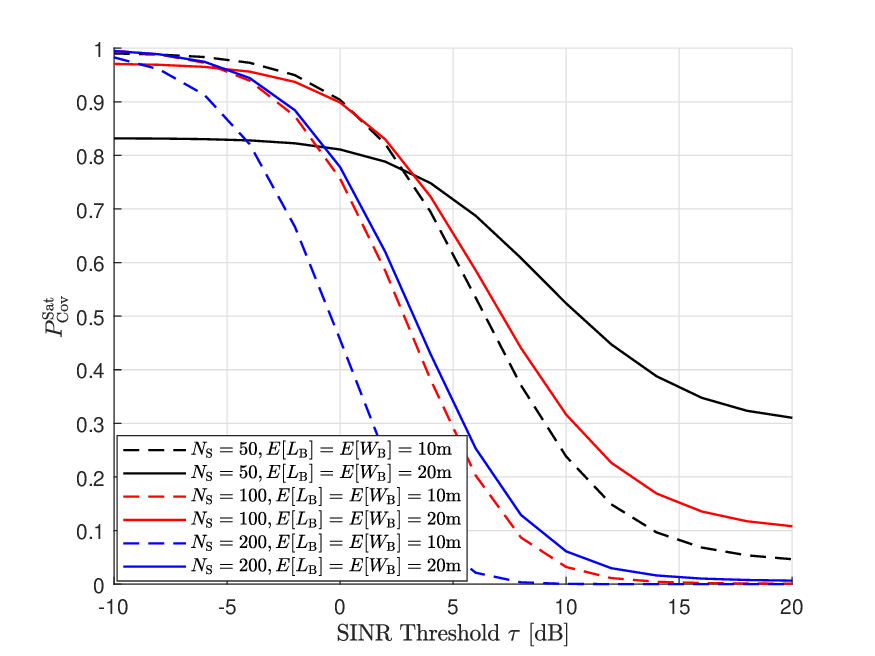}
    \label{fig2b}
}
\end{minipage}
\caption{Coverage probability analysis for satellite network. (a) Coverage probabilities with respect to the blockage density for $E[L_\mathrm{B}]=E[W_\mathrm{B}]=20\mathrm{m}$ and threshold SINR $\tau=0~\mathrm{dB}$. (b) Coverage probabilities with respect to SINR threshold $\tau$ when $E[h_\mathrm{B}]=50\mathrm{m}$ and blockage density $\lambda_\mathrm{B}=1000~\mathrm{blockages}/\mathrm{km}^2$.}
\vspace{-5mm}
\label{fig2}
\end{figure}

We first investigate the effect of the blockages on the satellite network. Fig.~2(a) illustrates the impact of the blockage density and the average blockage height on the coverage probability of the satellite network, where the SINR threshold is fixed at $\tau=0~\mathrm{dB}$ and the blockage footprint size is set to $E[L_\mathrm{B}]=E[W_\mathrm{B}]=20~\mathrm{m}$. Across all six configurations, blockages exhibit both beneficial and detrimental effects on coverage. When the blockage density is very small, the aggregate interference is strong, resulting in a low coverage probability. As the blockage density increases, blockages begin to obstruct interfering satellite links, which reduces the aggregate interference and improves coverage. However, when the blockage density becomes extremely large, the desired signal link is also likely to be blocked, causing the coverage probability to degrade. These results demonstrate that blockages can be beneficial for coverage probability due to their effectiveness in mitigating interference. Moreover, for a fixed average blockage height, the optimal blockage density increases with the number of visible satellites $N_\mathrm{S}$. This is because interference becomes more severe in denser satellite networks, requiring a higher blockage density to sufficiently suppress it. Conversely, for a fixed number of satellites, the optimal blockage density decreases as the average blockage height increases, since taller blockages are more effective at attenuating interference, thereby requiring a smaller density.

Fig.~2(b) illustrates the satellite coverage probability versus the SINR threshold $\tau$ under different blockage footprint sizes, with $\lambda_\mathrm{B}=1000$~blockages/$\mathrm{km}^2$, and $E[h_\mathrm{B}]=50~\mathrm{m}$. A key observation is that the beneficial blockage size depends on the SINR threshold. At low $\tau$, coverage is primarily limited by LoS availability, where smaller blockages improve coverage by increasing LoS probability of the serving link, whereas larger blockages lower the LoS probability, thereby increasing outages due to LoS unavailability and degrading coverage. In contrast, at high $\tau$, coverage becomes interference-limited, and larger blockages can be beneficial by blocking interfering satellite links, leading to a threshold-dependent crossover in the preferred blockage size.
\begin{figure}[!t]
\centering
\begin{minipage}[t]{0.49\linewidth}
\centering
\subfloat[]{
    \includegraphics[width=\linewidth]{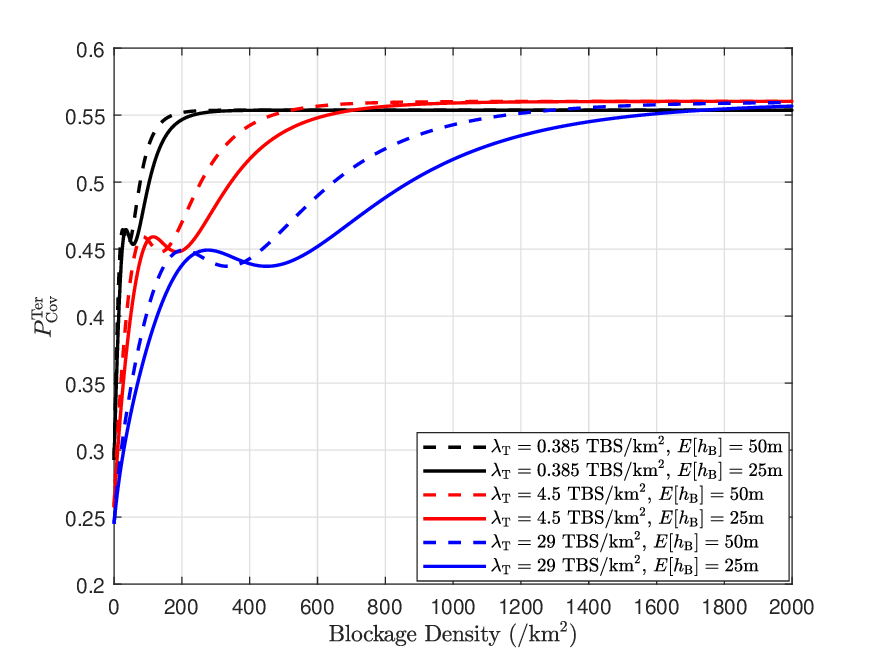}
    \label{fig3a}
}
\end{minipage}
\hfill
\begin{minipage}[t]{0.49\linewidth}
\centering
\subfloat[]{
    \includegraphics[width=\linewidth]{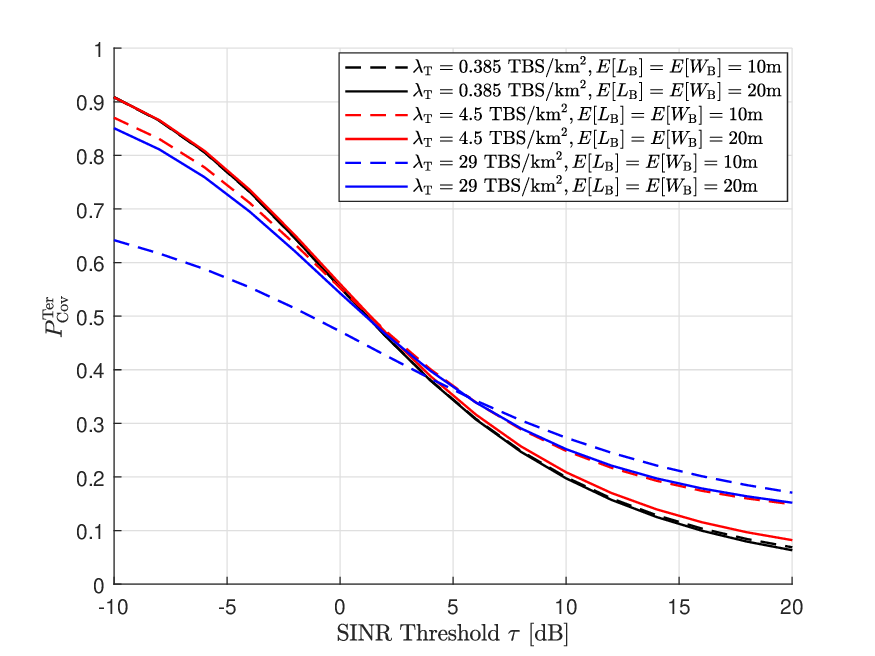}
    \label{fig3b}
}
\end{minipage}
\caption{Coverage probability analysis for terrestrial network. (a) Coverage probabilities with respect to the blockage density for $E[L_\mathrm{B}]=E[W_\mathrm{B}]=20\mathrm{m}$ and threshold SINR $\tau=0~\mathrm{dB}$. (b) Coverage probabilities versus SINR threshold $\tau$ when $E[h_\mathrm{B}]=50\mathrm{m}$ and blockage density $\lambda_\mathrm{B}$=1000~blockages$/\mathrm{km}^2$.}
\label{fig3}
\vspace{-5mm}
\end{figure}

Next, we examine the effect of blockages on the terrestrial network. Fig.~3(a) depicts the coverage probability for different deployment scenarios, including a rural environment with $0.385$ TBS$/\mathrm{km}^2$, a suburban environment with $4.5$ TBS$/\mathrm{km}^2$, and an urban environment with $29$ TBS$/\mathrm{km}^2$ \cite{3GPP1}. When the blockage density is low, most TBSs exhibit a high LoS probability; many interfering links remain in LoS, resulting in strong aggregate interference. As the blockage density increases, coverage typically exhibits three regimes: it first improves due to interference suppression, then decreases as the serving link is likely to become NLoS, and finally rebounds when interference mitigation dominates. Beyond a certain blockage density, the coverage saturates as the residual aggregate interference becomes negligible and the network becomes approximately noise-limited.

Fig.~3(b) illustrates the terrestrial coverage probability depending on the SINR threshold $\tau$ under various TBS densities and blockage footprint sizes. Unlike the satellite network, smaller blockages become advantageous at high SINR thresholds, because the coverage is dominated by the signal power of the serving link: larger blockages reduce the LoS probability of the serving link and thus attenuate the desired signal, while the corresponding interference reduction is insufficient to compensate under stringent SINR targets, making the benefit of smaller blockages more pronounced as $\tau$ increases.
\begin{figure} [!t]
\centering
\includegraphics[width=\columnwidth]{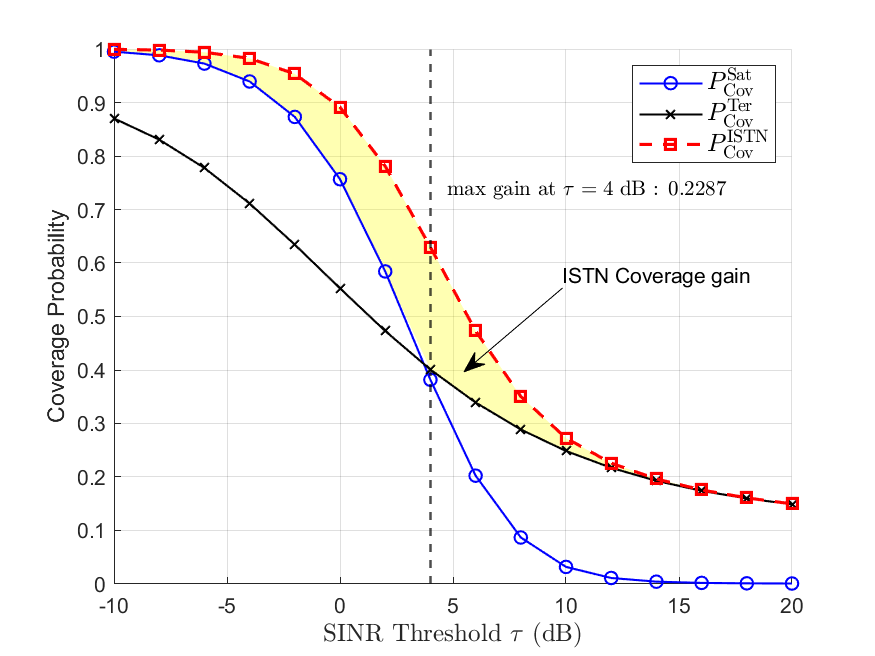}
\caption{Coverage probability of the ISTN, satellite network and terrestrial network. The shaded region represents the region where the usage of ISTN can produce a coverage gain.}
\label{fig4}
\vspace{-5mm}
\end{figure}

Fig.~4 illustrates the coverage improvement achieved by the ISTN. We consider an ISTN environment with $\lambda_\mathrm{B} = 1000$ blockages$/\mathrm{km}^2$, $4.5$ TBS$/\mathrm{km}^2$, and $N_\mathrm{S} =100$. The blockage parameters are set to $E[h_\mathrm{B}]=50~\mathrm{m}$ and $E[L_\mathrm{B}]=E[W_\mathrm{B}]=10~\mathrm{m}$. At very low $\tau$, the satellite network achieves near-unit coverage, and the resulting ISTN coverage gain is marginal. Similarly, at very high $\tau$, as the terrestrial network yields almost zero coverage, leading again to negligible gains. Consequently, the ISTN gain is maximized in an intermediate regime, where both individual networks provide only mediocre coverage. As shown in Fig.~4, integration provides an additional association path, widening the gap between $P_\mathrm{Cov}^\mathrm{ISTN}$ and the lower bound $\max(P_\mathrm{Cov}^\mathrm{Sat},P_\mathrm{Cov}^\mathrm{Ter})$. This indicates that the primary value of ISTN lies in stabilizing coverage probability over moderate SINR targets.

\section{Conclusion}
In this paper, we developed a stochastic geometry-based analytical framework to evaluate the downlink coverage probability of integrated satellite-terrestrial networks in the presence of terrestrial blockages. Unlike conventional models, the proposed analysis explicitly incorporates the effects of Earth's curvature and blockage geometry on low Earth orbit (LEO) satellite links. Our results reveal that terrestrial blockages play a dual role in network performance: while they attenuate desired signal links, they also effectively suppress aggregate interference. As a consequence, a moderate density of blockages can improve coverage probability in interference-limited scenarios. Furthermore, numerical results demonstrate that the proposed ISTN architecture provides meaningful coverage gains compared to terrestrial-only or satellite-only networks, particularly in challenging environments characterized by severe blockage or strong interference. These findings highlight the importance of blockage-aware modeling and joint satellite-terrestrial design for achieving reliable and ubiquitous connectivity in future 6G systems. As a promising direction for future work, the proposed framework can be extended to models that explicitly capture spatially correlated blockages.

\section*{Acknowledgment}
This work was supported by Institute of Information~\&~communications Technology Planning~\&~Evaluation (IITP) grant funded by the Korea government (MSIT) (No. RS-2021-ll210260, Research on LEO Inter-Satellite Links)

\end{document}